# Funding Covid-19 research:

*Insights from an exploratory analysis using open data infrastructures*

Alexis-Michel Mugabushaka (https://orcid.org/0000-0003-4624-568X)[1]
Nees Jan van Eck (https://orcid.org/0000-0001-8448-4521)[2]
Ludo Waltman (https://orcid.org/0000-0001-8249-1752)[2]

[1] European Commission, DG RTD, Unit G2[1]
[2] Centre for Science and Technology Studies (CWTS), Leiden University, The Netherlands

**Abstract**

**To analyse the outcomes of the funding they provide, it is essential for funding agencies to be able to trace the publications resulting from their funding. We study the open availability of funding data in Crossref, focusing on funding data for publications that report research related to Covid-19. We also present a comparison with the funding data available in two proprietary bibliometric databases: Scopus and Web of Science. Our analysis reveals a limited coverage of funding data in Crossref. It also shows problems related to the quality of funding data, especially in Scopus. We offer recommendations for improving the open availability of funding data in Crossref.**

## 1. Introduction

The on-going coronavirus 2019 pandemic (Covid-19) has caused a major public crisis. It has advanced to a major cause of death and overwhelmed healthcare systems in many countries. According to the WHO Covid-19 dashboard, as of January 2022, there have been over 340 million confirmed cases and 5.5 million deaths worldwide. The measures taken to contain its spread have also caused unprecedented disruptions of economic and social life around the world.

Researchers have been among the "first responding" professions dealing with the pandemic and its consequences. They advise public authorities on best measures to control the pandemic, study the course of the disease, develop clinical guidelines and medical protocols, and very importantly develop vaccines - some of which were developed in record time (Thomas et al., 2021) - as well as therapies.

Around the world, several research teams have redirected their research efforts to help fight the pandemic (Hao, 2020; Kwon, 2020; Viglione, 2020). Research funding bodies have multiplied initiatives to support research related to the pandemic. In addition to measures allowing grant management flexibility (Stoye, 2020), several organisations have launched fast track research funding programs specifically targeted at various aspects of the crisis. For example, in the US, the NIH launched several initiatives to tackle the pandemic by using existing funding mechanisms or establishing new, dedicated programs. They are bundled in the NIH-Wide Strategic Plan for COVID-19 Research. The NSF activated its Rapid Response Research mechanisms (RAPID) used for research funding in unanticipated events. In Europe, the European Union launched a Covid-19 emergency call for proposals already in January 2020 and published subsequent calls throughout the year. The Innovative Medicines Initiative (IMI) launched a fast-track call for proposals to speed up the development of new drugs and diagnostics to halt the global outbreak of Covid-19. The section "national activities" of the European Research Area (ERA) corona platform lists several other initiatives launched at the beginning of the pandemic[2]. They include the Deutsche Forschungsgemeinschaft (DFG - German Research Foundation), which set up a Covid-19 focused funding program, and the Swedish Research Council, which, among other initiatives, teamed up with the National Natural Science Foundation of China (NSFC) to support collaborative projects on coronavirus.

---

[1] The work was performed while AMM was on secondment to the European Commission (EC). The views expressed in this paper are the authors'. They do not reflect the views or official positions of the EC.

The OECD's COVID-19 WATCH, which monitors research policy responses to the Covid-19 crisis, estimates the combined value of public research funding in those measures to be about 2.6 billion dollars and 3.8 billion dollars if also other sources (charities, industry) are considered.

This has led to a massive expansion of Covid-19 research. Scientific publishers have adapted their editorial processes to allow fast dissemination of new results (Hurst & Greaves, 2021), posing to researchers and the public a particular "*challenge of discerning signal amidst noises*" as the editors of one journal put it (Kaplan et al., 2020).

The resulting unprecedented increase of research papers on a single topic - to some account over 100k in 2020 alone, accounting for about 4% of total research outputs (Else, 2020) - has also triggered a large body of meta-research on Covid-19 research. One strand of this research seeks to tame what has been termed a *"paper tsunami*" (Brainard, 2020). It uses advanced machine learning techniques for information extraction, misinformation detection, question answering etc. (Shorten et al., 2021). Another line of work uses scientometric techniques to develop an understanding of the output of Covid-19 publications. This line of research for instance studies the role of countries, institutions, journals and authors (Tao et al., 2020; El Mohadab et al., 2020), specific fields and techniques (Aristovnik et al., 2020; Hossain et al., 2020), gender (Andersen et al., 2020), research areas (Colavizza et al., 2021) and researchers (Ioannidis et al., 2021).

One aspect which remains under-explored is how this research has been, and is being, funded. One of the notable exceptions is the recent work by Cross et al. (2021) titled "*Who funded the research behind the Oxford-AstraZeneca Covid-19 vaccine?*". In this work, the authors analysed funding information of about 100 peer-reviewed articles relevant to the Chimpanzee adenovirus-vectored vaccine (ChAdOx) on which the Oxford-AstraZeneca vaccine is based. The authors found that this research was almost entirely supported by public funding. The European Commission, the Wellcome Trust, and the Coalition for Epidemic Preparedness Innovations (CEPI) were the biggest funders of ChAdOx research and development. The study also highlights the lack of transparency in reporting of funding, which "*hinders the discourse surrounding public and private contributions towards R&D and the cost of R&D.*"

Another study in this context is the analysis of how NIH funding has contributed to research used in Covid-19 vaccine development (Kiszewski et al., 2021). The authors focused on ten technologies employed in candidate vaccines (as of July 2020), identified from WHO documents, and on research on five viruses with epidemic potential. They then estimated the NIH funding to those areas by linking relevant publications (identified by searching in PubMed via MeSH terms) to grants using acknowledgments. The authors concluded that NIH funding has significantly contributed to advances in vaccines technologies, which helped the rapid development of Covid-19 vaccines. However, they also noted that NIH funding for vaccine research for specific pandemic threats has been inconsistent and called for sustained public sector funding for better preparedness against future pandemics.

---

[2] https://ec.europa.eu/info/funding-tenders/opportunities/portal/screen/covid-19?tabId=5

In this paper, we expand the scope from a single technology or single funding body to research on Covid-19 in general. Getting an accurate picture of Covid-19 research funding is important for a number of reasons:

- Insights into the various funding mechanisms and modalities used and how (relatively) successful they have been may inform the organisation of funding in case of future emergencies.
- Given the societal interest and policy implications of the outcomes of Covid-19 research, it is important to understand how these outcomes relate to the interests of sponsors. Most publication ethics guidelines require researchers to state the role of funders in reported research. While this is mostly applied in medical journals, extending it to research on other aspects of the pandemic can bring the transparency needed to assess the credibility of scientific findings.
- Most, if not all, public research organisations funding Covid-19 research have accountability obligations. They must report to public authorities on the results of their funding activities. Studying the funding patterns of this research can help funders understand not only the results of their activities but also how these results relate to research funded by others and, by putting it in a wider perspective, their relative weight in funding Covid-19 research.
- The concerns over fair vaccine access and the resulting debates on patent waivers for Covid-19 vaccines could be better informed by reliable data on the public investments which enabled the vaccine development.

In this paper, we explore the funding of Covid-19 research. The main objective is to find out which funding organisations have contributed to Covid-19 research reported in the scholarly literature. We seek specifically:

- to explore the extent to which funding data can be found in openly available databases, in particular Crossref;
- to identify the main funding organisations which supported Covid-19 research;
- to compare the findings based on openly available databases with those based on proprietary databases.

Another study of Covid-19 research funding was carried out by Shueb et al. (in press). Unlike our work reported in the present paper, the study by Shueb et al. covers only research published in the first months of the pandemic and does not make use of openly available funding data.

This paper is organised as follows. In Section 2, we discuss the data used in our analyses. We report our results based on openly available data in Section 3, while we present a comparison with results based on proprietary data in Section 4. We summarise our findings and draw conclusions in Section 5.

**2. Data**

We use the funding acknowledged by publications as a proxy for funding of the underlying research (Álvarez-Bornstein & Montesi, 2020). This requires combining data on Covid-19 related publications and data on the funding sources of these publications. In this section, we discuss the data we combined to link publications to funding as well as the data resulting from this linking.

*2.1. CORD-19 data*

We use the COVID-19 Open Research Dataset (CORD-19), a dataset of COVID-19 research articles (both metadata and full text) released by Semantic Scholar in partnership with other organisations. CORD-19 defines itself as "*a comprehensive collection of publications and preprints on COVID-19 and related historical coronaviruses such as SARS and MERS*" (Wang et al., 2020).

CORD-19 combines data from different sources which follow mainly the same search approach. It is updated regularly by adding new records and deleting erroneous or retracted entries. While CORD-19 is the most widely used Covid-19 literature dataset, it is not without limitations. The search approach used by CORD-19 has the advantage of conceptual clarity (i.e., papers included say something about the last three outbreaks caused by coronaviruses or about coronaviruses more generally), but this advantage is also its inherent limitation: keywords-based search may lead to the inclusion of papers which only cursorily mention a coronavirus outbreak (false positive) or miss relevant papers which use other terms (false negatives). Other limitations, acknowledged by the CORD-19 team, include the restriction of the data to scholarly publications, including pre-prints, leaving aside "*other types of documents that could be important, such as technical reports, white papers, informational publications by governmental bodies*" (Wang et al. 2020) as well as the focus on English language publications.

Some research has critically inspected the CORD-19 dataset with respect to coverage and has suggested possible improvements. Kanakia et al. (2020) explored how citation links can be used to understand and mitigate possible bias in CORD-19. Colavizza et al. (2020) also studied the coverage of CORD-19, using a version of the dataset from July 2020. Within the dataset, they identified a subset called "CORD-19 strict", for which the CORD-19 query matches the title and abstract of a publication, disregarding the full text. They found that this subset of CORD-19 almost perfectly matches the results retrieved from the Web of Science database, indicating that CORD-19 "*provides an almost complete coverage of research on COVID-19 and coronaviruses*". However, the fact that this subset is small suggests that CORD-19 does not only cover Covid-19 research, leading Colavizza et al. to caution users to be aware that CORD-19 may include "*a large number of publications whose relevance for COVID-19 and coronaviruses research needs a more careful assessment, and some of which may be of limited relevance*". The uncertainties in the scope of CORD-19 and inevitable errors due to its data collection approach are a limitation of our analysis which should be kept in mind when interpreting our results.

In the rest of this paper, we refer to publications in the CORD-19 dataset as Covid-19 publications. This should be understood as publications that are in a broad sense related to Covid-19, including publications that appeared before the start of the Covid-19 pandemic and that deal with coronavirus research more broadly.

We used the CORD-19 dataset released by the Allen Institute on 15 February 2021, in the version enriched by Microsoft Academic (MAG) - by adding publication identifiers from MAG.

*2.2. Crossref data*

Linking publications to funding sources is far from straightforward. As discussed elsewhere (Mugabushaka, 2020), several approaches can be used, each with their advantages and limitations. Our primary focus is on funding data provided by Crossref, although we also perform a comparison with funding data from proprietary databases.

*2.2.1. Crossref funding data*

Crossref is a not-for-profit organisation that provides an open infrastructure used by many stakeholders in the scholarly communication system. Its members include publishers, universities, pre-print services, and funding organisations. Its primary function is to enable parties globally to update and exchange metadata about the scholarly record, identified through Digital Object Identifiers (DOIs) and made open for all.

Crossref encourages its members to deposit metadata going beyond standard bibliographic information. Those rich metadata may also include funding data. According to the guidance Crossref gives to its members, funding data can be obtained from authors when they submit a manuscript or extracted from the acknowledgment or funding information section of a manuscript. As part of its data curation, Crossref can also add missing data, for example by inferring missing funder identifiers from funders names. The share of Crossref records with funding data has steadily increased to reach about 25% in 2019 (Hendricks et al., 2020, Figure 3).

Crossref makes data on funding together with other publication metadata openly available. We use Crossref's XML Metadata Plus Snapshot. The snapshot was downloaded on 5 March 2021. We consider only the 110,851,607 records classified as journal article, book content, conference paper, or pre-print[3].

*2.2.2. Crossref Funder Registry*

Funding data in Crossref is powered by a taxonomy of funders maintained in the Crossref Funder Registry. The Funder Registry was started by Elsevier in 2012 and was donated to Crossref. The curation of the registry is supported by Elsevier which reviews it every four to six weeks to add new funding entities as well as update or correct existing ones.

The Funder Registry assigns a unique identifier, a DOI, to each funder. The registry is organised in a hierarchy in which individual entries are linked to parent and child entries.

In Crossref, funding data may refer to any hierarchical level in the Funder Registry. To give two examples:

- In the case of the US National Institutes of Health (NIH), funding data may refer to a specific institute like the National Institute of Allergy and Infectious Diseases (DOI: 10.13039/100000060), the NIH as a whole (DOI: 10.13039/100000002), or the US Department of Health and Human Services (DOI: 10.13039/100000016).
- In the case of the European Union, funding data may refer to the Marie Skłodowska-Curieprogram (DOI: 10.13039/100010665), the European Research Council (DOI: 10.13039/100011199), the H2020 program (DOI: 10.13039/501100007601), or the European Commission (DOI: 10.13039/501100000780).

Funding data captured by publishers and submitted to Crossref reflects the different acknowledgment practices of authors. This can lead to an inconsistent picture of the contributions of funders of Covid-19 research. For a more accurate picture, there is a need to group funders based on the hierarchy of funding organisations in the Funder Registry. Because this hierarchy follows the legal structure of funding bodies, it can, in some cases, make comparisons difficult. For example, public funding bodies in Canada have the Government of Canada at the highest level in the hierarchy of the Funder Registry, while in the US government departments usually constitute the highest level. We created a mapping of each entry in the Funder Registry to the corresponding top-level entity in the Funder Registry hierarchy (Van Eck & Mugabushaka, 2021). The mapping is based on version 1.34 of the Funder Registry, which has 27,741 entries at lower levels, which are grouped into 22,369 funders at the highest level. While not perfect, this approach has the advantage of transparency and simplicity. Creating an alternative mapping would not only require detailed knowledge of funding structures worldwide, but would also require subjective choices which may bias our analysis.

---

[3] In late 2019, funding organisations started registering grant metadata in Crossref, which is then linked to publications. In this paper we focus on funding data submitted to Crossref by publishers.

*2.3. Linking the datasets*

Figure 1 illustrates how the datasets were linked. We used DOIs to link CORD-19 publications to publications in Crossref. The CORD-19 version we used includes 484,064 records, of which 474,691 are unique publication records (i.e., unique CORD-19 identifiers). Of these records, 260,636 (or about 55%) have a DOI in CORD-19. After eliminating duplicates, we ended up with 259,652 unique DOIs, of which 255,378 were found in Crossref. Our analysis is based on these 255,378 records in Crossref.

The lack of DOIs for a substantial share of the publications in CORD-19 - including about half of the publications in 2020 and 2021 - is another important limitation of our analysis. It means that the results reported in the subsequent sections offer only a partial picture of the funding of Covid-19 research.

Of the CORD-19 publications linked to Crossref, 44,820 have funding data. For 36,008 publications, we also have an identifier of a funding organisation included in the Crossref Funder Registry. Our analysis of Covid-19 funding is based on these 36,008 publications.

The relatively low share of publications with funding data in Crossref is another limitation of our analysis. An important implication of this limitation is the need to pay attention to the way in which funding data is collected, and which measures can be taken by various stakeholders to increase the availability of funding data in open data infrastructures. We will share some reflections and suggestions in the concluding section.

**Figure 1: Funding data for Covid-19 publications in Crossref**

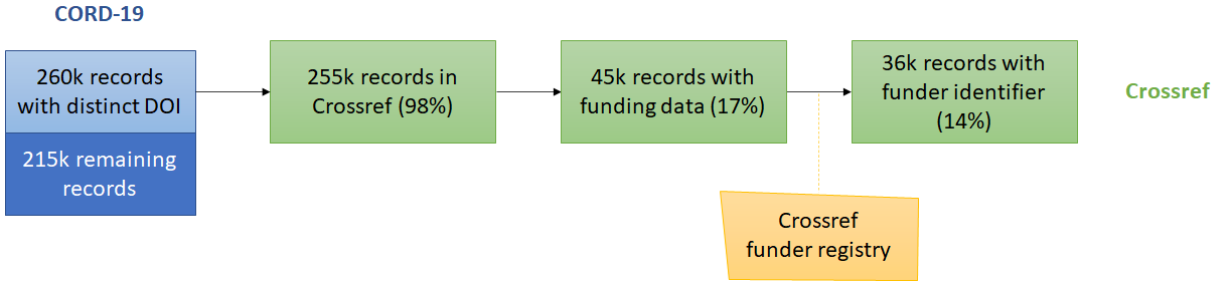

## 3. Analysis of open funding data in Crossref

The data at hand, effectively a relatively limited subset of Covid-19 research papers due to the data limitations described above, offers an incomplete picture of Covid-19 funding. As partial as the results are, however, they can give indications of funding patterns notably for the most prolific funders and how they interact in the network of "co-funding". The results also provide insight into the type of funding bodies supporting Covid-19 research.

*3.1. Availability of funding data*

As noted above, we focus on publications indexed in Crossref, as our aim is to use open funding data. Funding data is available in Crossref for 44,820 Covid-19 publications. This accounts for 17% of the Covid-19 publications for which a DOI is available.

As Figure 2 shows, the availability of funding data in Crossref has increased over time, reaching almost 40% in 2019. This is in line with earlier analyses showing that the amount of funding data submitted by publishers to Crossref has steadily increased (Hendricks et al., 2020; Van Eck & Waltman, 2021). Before 2020, the share of Covid-19 publications with funding data is higher than the overall share of publications with funding data. The sharp decline in 2020 in the share of publications with funding data may seem puzzling. It could be that in the beginning of the Covid-19 pandemic many researchers immediately started to work on Covid-19-related research projects, without first applying for funding. This impression is confirmed when looking at other databases. Both Web of Science and Scopus also show a significant decrease in the share of publications with funding data, reinforcing the idea that a relatively large share of all Covid-19 research in 2020 did not receive funding from a funding agency.

**Figure 2: Percentage of Covid-19 publications with a DOI that have funding data in Crossref**

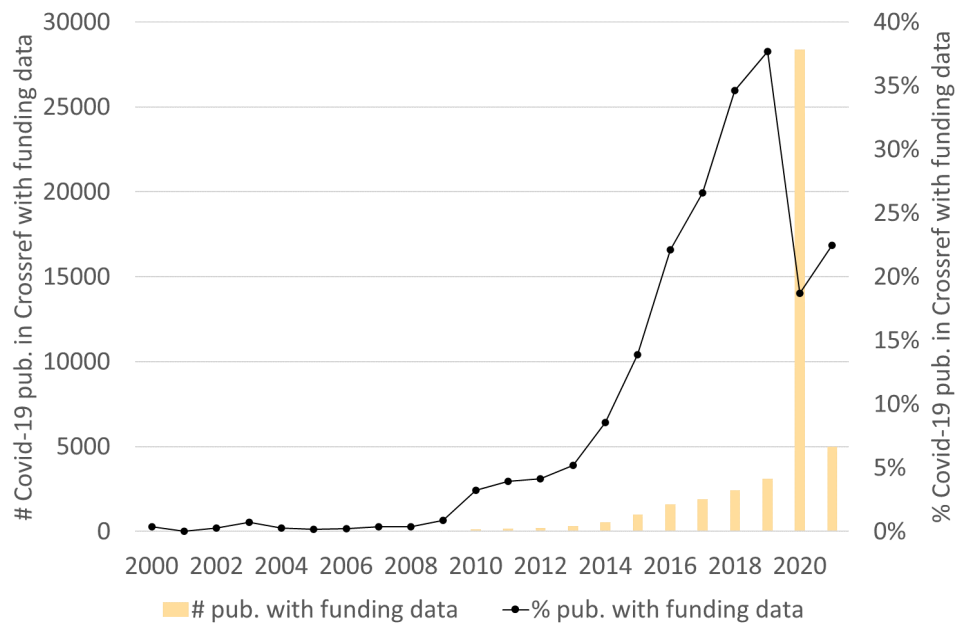

### 3.2. Top funders

The 36,008 publications linked to a funder identifier acknowledge 5,386 distinct funders at the top level of the Crossref Funder Registry hierarchy. This indicates that, based on open funding data in Crossref, close to one in four funding bodies at the top level in the Funder Registry have supported Covid-19 research. The number of papers per funder varies significantly. The median is two papers. The top 10 funders account for over half of all papers.

Table 1 shows the top 30 funders with the largest number of Covid-19 publications. Based on open funding data in Crossref, the top funders are the US Department of Health and Human Services with about 7.1k publications and the National Natural Science Foundation of China with 5.3k publications. They account for 20% and 15%, respectively, of the papers for which funding data is available. They are followed by the European Commission in the third place with 1.3k publications and by UK Research and Innovation and the Ministry of Science and Technology of the People's Republic of China, each with 1.1k publications. Other funders in the top 10 are the Government of Canada, the US National Science Foundation, the Deutsche Forschungsgemeinschaft, the Japanese Ministry of Education, Culture, Sports, Science and Technology, and the UK National Institute for Health Research.

**Table 1: Top 30 funders of Covid-19 publications (based on Crossref data)**

| Country | Funder | # pub. |
|---|---|---|
| USA | U.S. Department of Health and Human Services | 7,081 |
| CHN | National Natural Science Foundation of China | 5,318 |
|  | European Commission | 1,313 |
| CHN | Ministry of Science and Technology of the People's Republic of China | 1,128 |
| GBR | UK Research and Innovation | 1,111 |
| CAN | Government of Canada | 970 |
| USA | National Science Foundation | 942 |
| DEU | Deutsche Forschungsgemeinschaft | 730 |
| JPN | Ministry of Education, Culture, Sports, Science and Technology | 691 |
| GBR | National Institute for Health Research | 655 |
| BRA | Ministério da Ciência, Tecnologia e Inovação | 560 |
| KOR | National Research Foundation of Korea | 536 |
| USA | U.S. Department of Defense | 531 |
| AUS | Department of Health, Australian Government | 498 |
| USA | Bill and Melinda Gates Foundation | 473 |
| CHN | Ministry of Education of the People's Republic of China | 466 |
| BRA | Coordenação de Aperfeiçoamento de Pessoal de Nível Superior | 450 |
| ESP | Ministerio de Economía y Competitividad | 418 |
| CHN | Ministry of Finance | 343 |
| GBR | Wellcome Trust | 340 |
| CHN | China Postdoctoral Science Foundation | 298 |
| BRA | Fundação de Amparo à Pesquisa do Estado de São Paulo | 273 |
| FRA | Agence Nationale de la Recherche | 271 |
| CHE | Schweizerischer Nationalfonds zur Förderung der Wissenschaftlichen Forschung | 265 |
| TWN | Ministry of Science and Technology, Taiwan | 261 |
| JPN | Japan Agency for Medical Research and Development | 256 |
| USA | U.S. Department of Agriculture | 251 |
| IND | Department of Science and Technology, Ministry of Science and Technology, India | 236 |
| USA | Foundation for the National Institutes of Health | 229 |
| DEU | Bundesministerium für Bildung und Forschung | 223 |

*3.3. Type of funder*

For each funding body, the Crossref Funder Registry includes information about the type of organisation. This information is organised in two dimensions. On the one hand, a distinction is made between private and public organisations. On the other hand, for each of these categories, a further distinction is made between different organisation forms.

**Table 2: Number of Covid-19 publications by type of funding organisation (based on Crossref data)**

| Type of funding organisation | Public | Private | All |
|---|---|---|---|
| National government | 23,959 |  | 23,959 |
| Trusts, charities, foundations (both public and private) | 10 | 6,355 | 6,365 |
| Local government | 6,309 |  | 6,309 |
| Universities (academic only) | 843 | 3,732 | 4,575 |
| Other non-profit organisations | 18 | 3,073 | 3,091 |
| For-profit companies (industry) | 577 | 1,204 | 1,781 |
| Associations and societies (private and public) | 55 | 987 | 1,042 |
| International organisations | 6 | 742 | 748 |
| Research institutes and centers | 59 | 274 | 333 |
| Total | 31,836 | 16,367 | 48,203 |
| Distinct total | 28,186 | 13,900 | 42,086 |

Table 2 shows that over two-third of the publications (28k) acknowledge funding bodies classified as "government" while the rest (14k) acknowledge funding from private entities.

The classification of funding organisations in the Funder Registry follows the legal status of these organisations in different countries. Given their particularities, this can lead to results that are difficult to compare. For example, one of the major public funding bodies in Germany, the DFG, is classified in the Funder Registry as a private organisation under "trusts, charities, foundations", which is indeed its legal form. In a way, however, it is comparable to other funders classified as "government" – like the NIH and NSF in the US - as it receives its funding from public authorities (both at the federal and the local level).

The classification of organisations in the Funder Registry allows us to identify other non-public players active in funding Covid-19 research. Among those with more than 100 publications, we find philanthropic organisations like the Wellcome Trust and the Bill & Melinda Gates foundation and pharmaceutical companies like Pfizer, Sanofi, and Novartis (see 'Sec_3_2' in the supplementary material; Mugabushaka et al., 2022).

*3.4. Co-funding rate and co-funding network*

As shown in Figure 3, one-third of the Covid-19 papers with funding data are linked to more than one funding body. In these cases, the research team behind the reported work may have multiple lines of funding that contributed to the reported results, or the authors may belong to multiple research teams with different sources of funding.

**Figure 3: Number of Covid-19 publications by number of funders (based on Crossref data)**

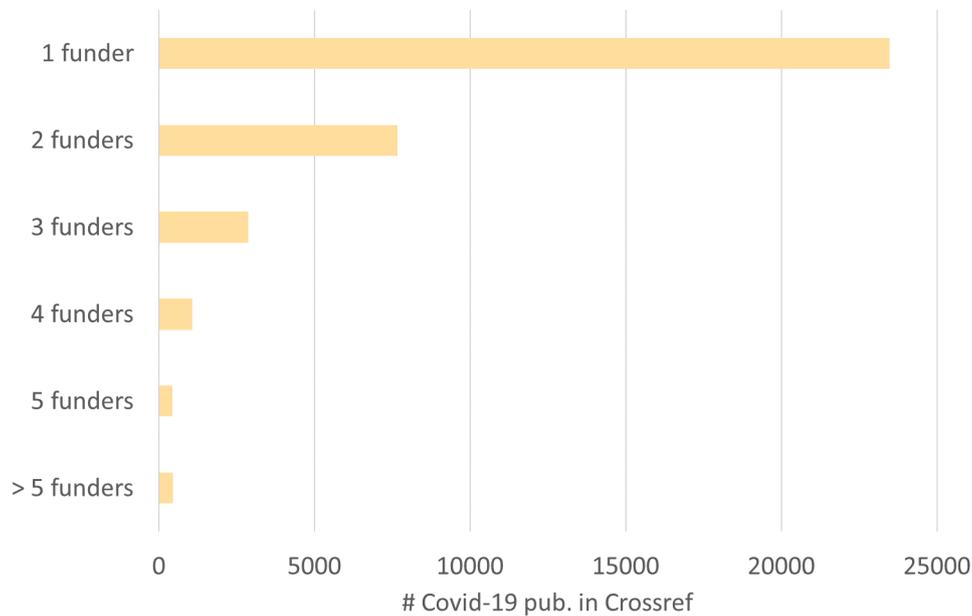

Through papers acknowledging multiple sources of funding, funding bodies are effectively engaged in a co-funding collaboration network. Looking at the funders with the largest number of publications (see 'Sec_3_4b' in the supplementary material; Mugabushaka et al., 2022), we see that the share of publications in which a funding agency is acknowledged together with other funders is, on average, relatively high, but with some variation across funders. For some of the large funders, the share of publications co-funded with other funders is around 50%. Examples are the National Natural Science Foundation of China (48%), the Government of Canada (55%), the Deutsche Forschungsgemeinschaft (55%), and the US Department of Health and Human Services (58%). For other large funders, such as UK Research and Innovation (68%) and the European Commission (72%), more than two-third of the publications are co-funded with other funders.

Figure 4 presents a visualisation of a co-funding network for Covid-19 publications. The visualization was created using the VOSviewer software (Van Eck & Waltman, 2010). The network includes 384 funders that each have at least 30 co-funding links with other funders in the network. The visualisation can be explored interactively at https://tinyurl.com/z27f97ek.

**Figure 4: Co-funding network for Covid-19 publications (based on Crossref data)**

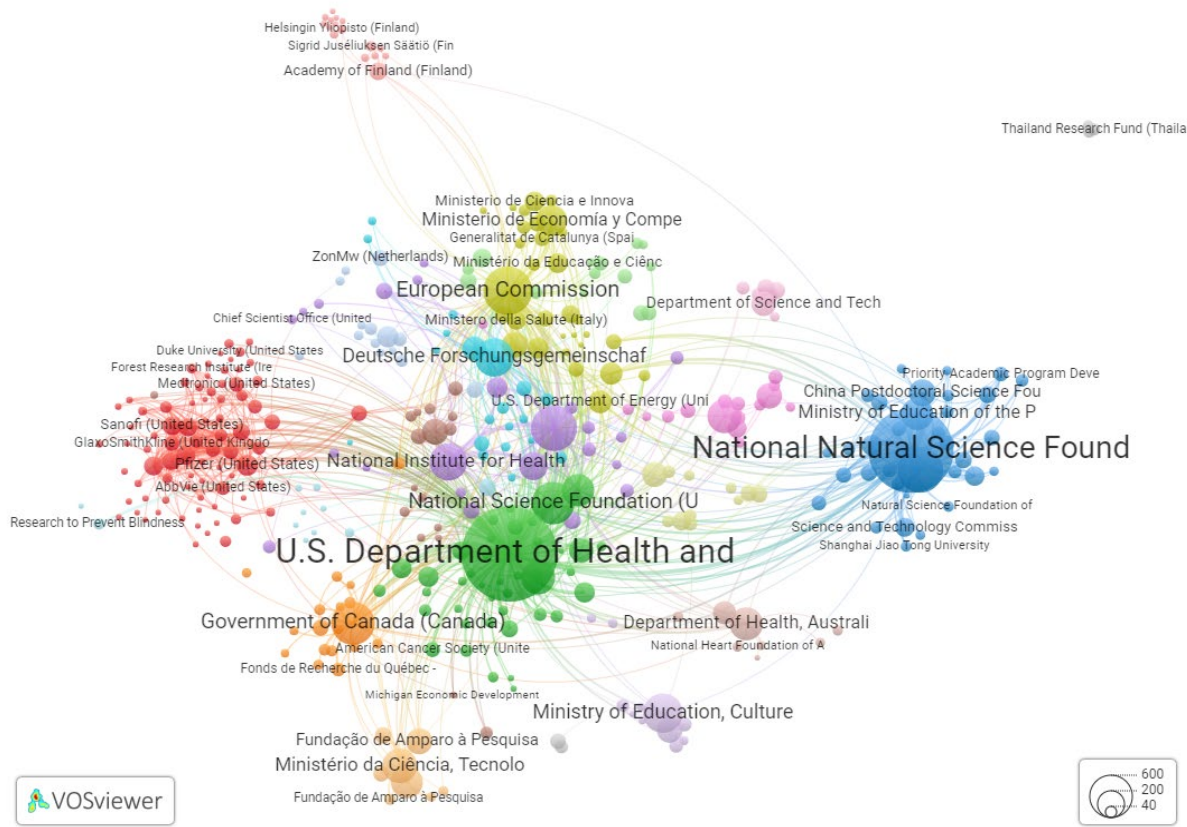

## 4. Comparison with proprietary funding data

To assess the comprehensiveness of open funding data in Crossref, we performed a comparison with funding data in two proprietary bibliometric databases: Scopus and Web of Science (WoS). Following the approach taken for Crossref, we use the subset of the CORD-19 dataset with DOIs to query Scopus and WoS and retrieve funding data. The comparison with funding data in proprietary databases aims to provide insight into the comprehensiveness of open funding data in Crossref. Comparing the funding data made available by different proprietary databases is not the main purpose of our analysis. For earlier analyses of funding data available in Scopus and WoS, we refer to Paul-Hus et al. (2016), Álvarez-Bornstein et al. (2017), Grassano et al. (2017), Tang et al. (2017), Kokol and Blažun Vošner (2018), Liu (2020), and Liu et al. (2020).

To have a meaningful comparison, we focus on funding data obtained from publishers and made available in bibliometric databases. We do not consider funding data collected from funding agencies. Considering data obtained from funding agencies would obscure the comparison between Crossref and proprietary bibliometric databases, since our analysis for Crossref takes into account only data obtained from publishers. WoS nowadays also includes funding data obtained from funding agencies, such as data from NIH Reporter, but we do not use this data. We also do not consider funding data from the Dimensions database (Herzog et al., 2020), since this database does not make a distinction between funding data obtained from publishers and from funding agencies.

The Scopus and WoS data were retrieved from the in-house database system of the Centre for Science and Technology Studies (CWTS) at Leiden University. For both databases, we used data from April 2021. The following WoS citation indexes were used: Science Citation Index Expanded (SCIE), Social Sciences Citation Index (SSCI), Arts & Humanities Citation Index (AHCI), and Conference Proceedings Citation Index (CPCI). We did not use the Emerging Sources Citation Index, because this citation index is not included in the WoS license of CWTS.

Scopus uses the same funder registry as Crossref, making it relatively easy to compare the funding data available in Crossref and Scopus. WoS takes a different approach and uses its own funder registry. This registry provides a unified name to each funder. Due to the different approach taken by WoS, there is no easy way to compare Crossref and WoS in terms of the availability of funding data at the level of individual funders. We therefore compare Crossref and WoS only in terms of whether publications do or do not report funding, without taking into account the funder that provided the funding.

In this section, we first analyse the availability of funding data in Scopus and WoS, we then look at the top funders, and finally we explore the differences between Scopus, WoS, and Crossref.

*4.1. Availability of funding data*

Bibliometric databases have different scopes due to differences in their inclusion criteria. Visser et al. (2021) recently reported that overall WoS covers fewer publications than Scopus, even though there are quite some publications, for instance meeting abstracts and book reviews, that are covered by WoS and not by Scopus. For DOIs in CORD-19, Figure 5 shows that Scopus indexes more publications than WoS (187,518 vs. 171,130). However, looking at publications with funding data, Scopus has a lower coverage than WoS (61,168 vs. 73,444 publications). On the other hand, Scopus has a higher coverage than WoS (52,747 vs. 46,070 publications) if we consider only publications that include funder identifiers or unified funder names.

Both in Scopus and in WoS, the availability of funding data is higher than in Crossref, where funding data is available for 44,820 publications, of which 36,008 include funder identifiers (see Figure 1).

**Figure 5: Funding data for Covid-19 publications in Scopus and WoS**

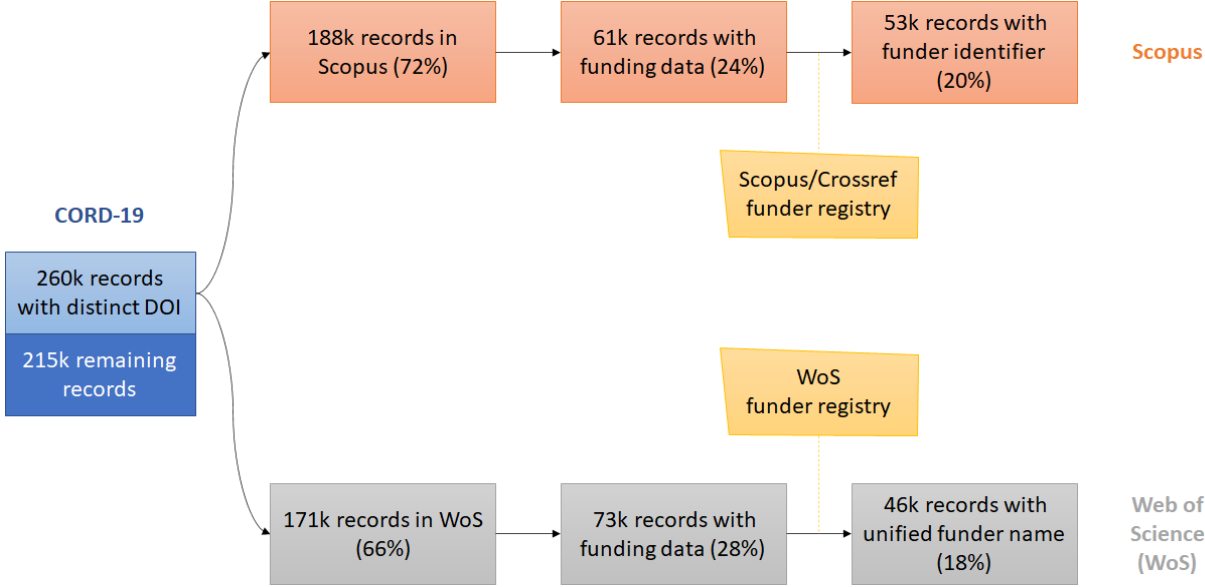

*4.2. Top funders*

Scopus uses the Crossref Funder Registry described in Section 2.2.2, while WoS has its own registry of funders. On account of this, we do not make a direct comparison of the number of publications per funder in Scopus and WoS. Instead, we present separate statistics for each of the two databases.

For Scopus and WoS, Table 3 shows the top funders in terms of the number of Covid-19 publications. For Scopus, we look at funders at the highest level of the Funder Registry hierarchy, following the approach discussed in Section 2.2.2. For WoS, we use unified funder names.

Table 3 provides three interesting insights:

- First, the table shows that by and large, the top funders are the same in Scopus and WoS and match those in the open data obtained from Crossref. In fact, the first three funders are the same across the three databases: The US Department of Health & Human Services, followed by the National Natural Science Foundation of the China (NSFC) and the European Commission. Other top funders listed in Table 1 based on Crossref are also visible among the top funders listed in Table 3 based on the proprietary databases.
- A second observation is the difficulty to make comparisons between databases that use different registries of funders. WoS harmonises funder names, but unlike Scopus and Crossref, it does not enable funders to be aggregated into higher-level entities. While one may intuitively infer that the different institutes of the NIH belong to the same higher-level entity, it requires a considerable knowledge of the funding landscape to know that the European Research Council, which is listed as a separate organisation in the case of WoS, is part of the European Commission in the case of Scopus and Crossref. Hence, when comparing funding data from different databases, it is essential to pay close attention to the way in which relations between funding entities are handled.
- A third observation relates to pharmaceutical companies, which feature prominently on the Scopus list, but not on the Crossref and WoS lists.

We explore the differences between the three databases in more detailed in the next section.

**Table 3: Top 30 funders of Covid-19 publications (based on Scopus and WoS data)**

| Scopus | | WoS | |
|---|---|---|---|
| Funder | # pub. | Funder | # pub. |
| U.S. Department of Health and Human Services | 13,559 | United States Department of Health & Human Services | 11,857 |
| National Natural Science Foundation of China | 5,938 | National Institutes of Health (NIH) - USA | 11,341 |
| European Commission | 3,373 | National Natural Science Foundation of China (NSFC) | 7,946 |
| Ministry of Science and Technology of the People's Republic of China | 2,380 | European Commission | 4,875 |
| UK Research and Innovation | 2,269 | NIH National Institute of Allergy & Infectious Diseases (NIAID) | 1,832 |
| National Science Foundation (US) | 1,587 | National Science Foundation (NSF) | 1,464 |
| Ministry of Education, Culture, Sports, Science and Technology (Japan) | 1,585 | German Research Foundation (DFG) | 1,424 |
| Government of Canada | 1,454 | Canadian Institutes of Health Research (CIHR) | 1,305 |
| Deutsche Forschungsgemeinschaft | 1,146 | Ministry of Education, Culture, Sports, Science and Technology, Japan (MEXT) | 1,284 |
| National Institute for Health Research (UK ) | 1,118 | Medical Research Council UK (MRC) | 1,254 |
| Wellcome Trust | 1,022 | UK Research & Innovation (UKRI) | 1,221 |
| Ministry of Education of the People's Republic of China | 972 | Wellcome Trust | 1,201 |
| Pfizer | 866 | National Council for Scientific and Technological Development (CNPq) | 1,022 |
| Ministério da Ciência, Tecnologia e Inovação (Brazil) | 819 | National Health and Medical Research Council of Australia | 1,012 |
| U.S. Department of Defense | 798 | Japan Society for the Promotion of Science | 901 |
| Department of Health, Australian Government (Australia) | 748 | CAPES (Brazil) | 740 |
| National Research Foundation of Korea | 697 | Fundamental Research Funds for the Central Universities | 739 |
| Ministerio de Economía y Competitividad (Spain) | 658 | National Institute for Health Research (NIHR) | 736 |
| Ministry of Finance (China) | 633 | NIH National Cancer Institute (NCI) | 722 |
| Coordenação de Aperfeiçoamento de Pessoal de Nível Superior (Brazil) | 609 | NIH National Heart Lung & Blood Institute (NHLBI) | 685 |
| Merck | 582 | French National Research Agency (ANR) | 668 |
| Bill and Melinda Gates Foundation | 580 | Natural Sciences and Engineering Research Council of Canada (NSERC) | 641 |
| AstraZeneca | 577 | United States Department of Defense | 622 |
| Novartis | 576 | Bill & Melinda Gates Foundation | 620 |
| Roche | 567 | European Research Council (ERC) | 578 |
| Covidien | 551 | National Basic Research Program of China | 560 |
| Auris Health | 549 | CGIAR | 555 |
| Schweizerischer Nationalfonds zur Förderung der Wissenschaftlichen Forschung | 532 | NIH National Institute of General Medical Sciences (NIGMS) | 552 |
| Chinese Academy of Sciences | 505 | Swiss National Science Foundation (SNSF) | 538 |
| Medtronic | 483 | Ministry of Science and Technology, Taiwan | 533 |

*4.3. Differences between the databases*

In the previous sections, we analysed differences at an aggregate level in funding data obtained from three different databases: Crossref, Scopus, and WoS. In this section, we present an analysis at a more detailed level, focusing on the extent to which – at the level of individual publications – funding data obtained from these three databases differs or overlaps.

*4.3.1. Intersections and differences*

Figure 6 shows the overlap and the differences between the three databases in terms of publications for which funding data is available. The relatively small overlap is remarkable. There are 95,292 publications with funding data in at least one of the three databases. Only 23,950 of these publications have funding data in all three databases, an overlap of 25%. The number of publications with funding data in only one of the databases is largest for WoS (16k). It is somewhat smaller for Scopus (13k) and smallest for Crossref (6k).

**Figure 6: Overlap of Crossref, Scopus, and WoS in terms of Covid-19 publications with funding data (considering all publications indexed in at least one of the three databases)**

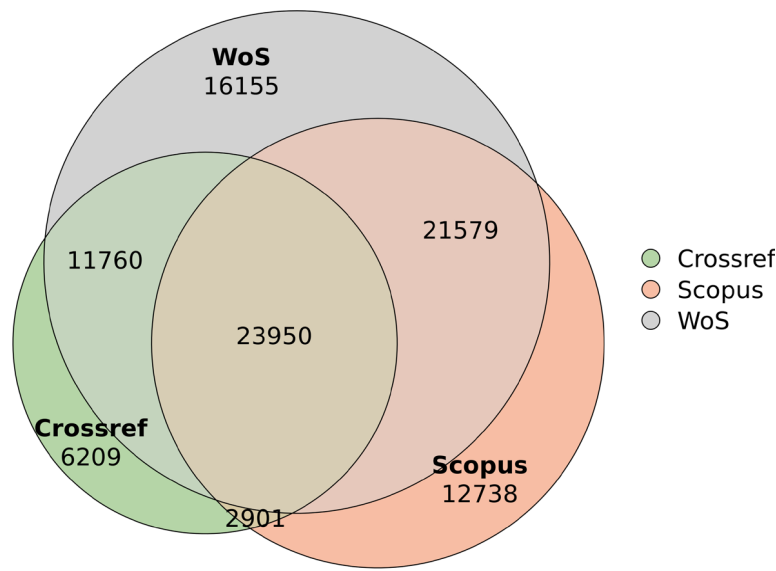

The differences shown in Figure 6 are partly due to differences in the publications indexed in the three databases. As indicated in Figure 5, of the DOIs in CORD-19, only 72% can be linked to publications indexed in Scopus and only 66% to publications indexed in WoS. In contrast, 98% of the DOIs can be linked to publications in Crossref, as shown in Figure 1.

In Figure 7, we therefore restrict the analysis to the 141,291 publications indexed in all three databases. Of these publications, 72,402 have funding data in at least one of the databases and 23,950 have funding data in all three databases, resulting in an overlap of 33%. Like in Figure 6, the number of publications that have funding data in one database but not in the others is largest for WoS (12k). It is somewhat smaller for Scopus (8k) and smallest for Crossref (1k).

**Figure 7: Overlap of Crossref, Scopus, and WoS in terms of Covid-19 publications with funding data (considering only publications indexed in all three databases)**

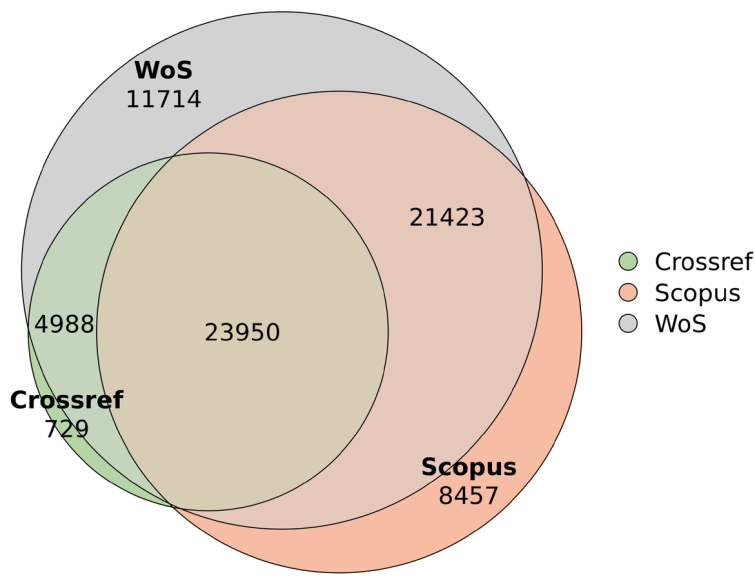

*4.3.2. Accuracy of funding data*

We now analyse the accuracy of funding data for individual publications by comparing funding data obtained from the different databases with funding information found in the full text of publications. The comparison is based on a stratified random sample of 120 publications. For each of the three databases considered, we drew a random sample of 40 publications that have funding data only in that database and not in the other two databases. In addition, given the notable presence of pharmaceutical companies among the funders of Covid-19 publications in Scopus, we also analyse a random sample of 25 publications that have at least one private entity as funder in Scopus (i.e., an entity classified in the Funder Registry as "private" and "for-profit company/industry"). The samples are available in the supplementary material (Mugabushaka et al., 2022).

Our findings can be summarized as follows:

- In the *Crossref sample*, we found the funding data for 27 of the 40 publications to correspond to the text in either the acknowledgment or the funding information section of a publication. In the following we refer to this as "correct" funding data. The correct entries include 2 instances in which the funding information is in fact a statement that there was no specific funding supporting the work (e.g., *"This work received no specific grant from a funding agency"*). Although the sample is too small to generalize to Crossref as a whole, this type of funding statement may be an important one that deserves further analysis. Databases that provide funding data may consider including it in their taxonomies. Of the 13 publications with incorrect funding data, 4 were apparently due to an error of the extraction algorithm, which for example mistook the affiliation of the authors for a funding body. In 1 case, the funding information was partially correct: one funding organization listed in the acknowledgment was missed but others correctly identified. In the other 8 cases, the funding information could not be located anywhere in the full text of the publication.

- In the *Scopus sample*, for 15 of the 40 publications, the funding data corresponds to the funding statement found in the full text. 25 cases were found to be errors, most probably of the algorithm for extracting funding information. The most common error was the algorithm incorrectly identifying the section of a publication that includes a funding statement. Sometimes a conflict-of-interest section was incorrectly interpreted as a funding statement. In other cases, the acknowledgement section was interpreted as a funding statement, while the publication in fact included a separate funding information section. Other errors for example included mistaking a natural person thanked by the authors for a funding body, or interpreting the affiliation of a researcher mentioned in the acknowledgment section as a funding body. In 4 cases, the funding information could not be found anywhere in the full text of the publication[4].
- In the *WoS sample*, we found that in 37 of the 40 cases the funding data corresponds to the funding statement in the full text. In 2 of these cases, we noted that the funding information provided in the full text was ambiguous. The funding information section stated that there was no funding to report, but the acknowledgment section mentioned a funding body that provided financial support. In the 3 cases in which no relevant funding statement could be found in the full text, there was an error, most probably caused by the extraction algorithm mistaking a conflict-of-interest section for a funding statement (see also Grassano et al., 2017; Lewison & Sullivan, 2015).
- In the *sample of publications with Scopus funding data that includes a private entity*, 5 of the 25 publications indeed contained a funding statement mentioning the private entity. In the other 20 cases, the funding data was incorrect: The private entity was mentioned in the conflict-of-interest section, but not as a funder of the research. In Scopus, this problem occurred in all 20 publications. It also occurred in 4 publications in WoS and in 1 publication in Crossref.

*4.3.3. Differences by publisher*

We now turn to differences by publisher in the coverage of funding data. Figure 8 includes – for each of the databases considered - the share of publications with funding data. We restricted the analysis to publications from 2020 and 2021 and show only publishers with 500 or more Covid-19 publications. Statistics for other publishers are available in the supplementary material (Mugabushaka et al., 2022).

Figure 8 shows some interesting differences between Scopus and WoS, but these are not as striking as the differences between the two proprietary databases and Crossref.

Two publishers, Oxford University Press and American Chemical Society, do an excellent job in submitting funding data to Crossref. For these publishers, the number of publications with funding data in Crossref is almost as large as in WoS and, in the case of Oxford University Press, substantially larger than in Scopus.

For many other publishers, the number of publications with funding data in Crossref is considerably below the corresponding number in WoS, and in most cases also below the corresponding number in Scopus. These publishers seem to have gaps in the funding data they submit to Crossref, or they may have started submitting funding data only recently.

Figure 8 also reveals three publishers which do not submit funding data to Crossref at all: American Medical Association, Cambridge University Press, and JMIR.

---

[4] Scopus informed us that its funding data should be seen as "*work in progress*" as there are still many incremental improvements in the planning. Currently Scopus focuses on optimizing the data for the top 300 funders.

There is a need to better understand why some publishers include funding information in the metadata they provide to Crossref while others do not. A possible explanation is that awareness among publishers of the importance of submitting funding data to Crossref still needs to grow, and some publishers may also need more time to implement the submission of funding data throughout all their workflows.

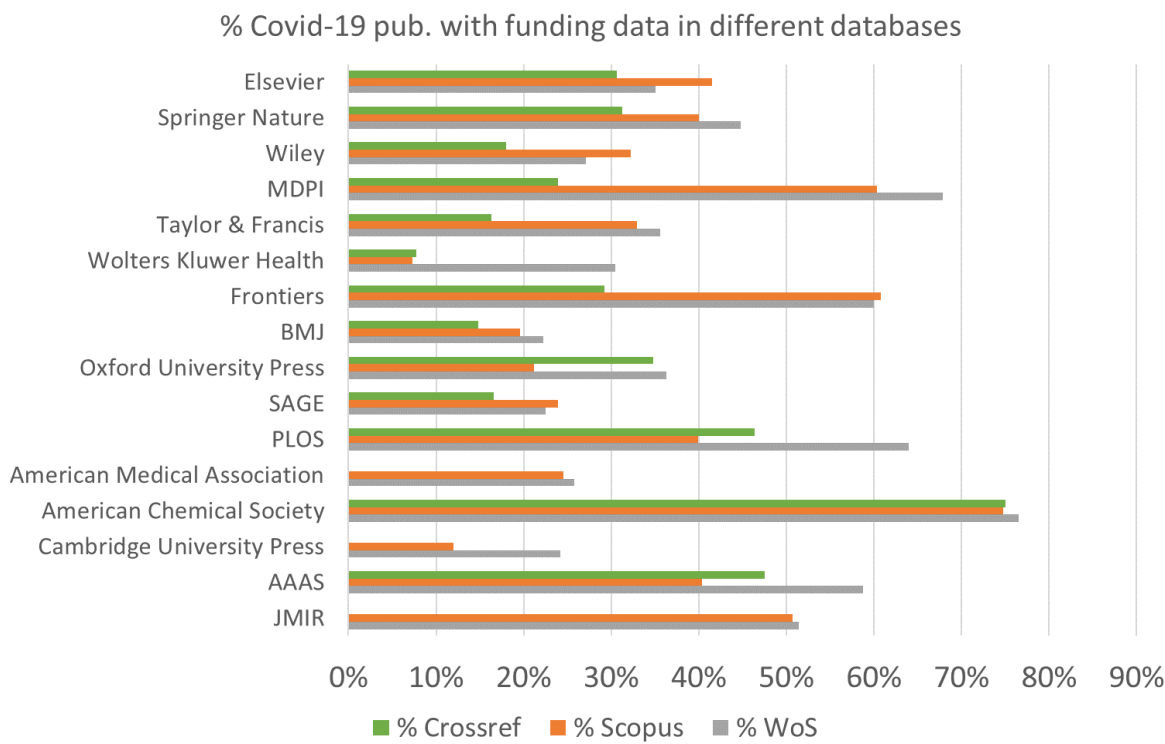

**Figure 8: Percentage of Covid-19 publications with funding data, breakdown by publisher and database (considering only publications indexed in all three databases)**

## 5. Summary and outlook

The coronavirus (Covid-19) pandemic has turned the world upside down. As the surge of cases threatens to overwhelm health care systems and death toll increases around the world, the effects of containment measures reverberate through economies and societies, sending shock waves that many fear will be also felt in years to come. Researchers have been among the first professions contributing to tackling the pandemic, and research funding organisations have adapted their programs or developed new ones to support them.

As a result, there has been an explosion of scientific papers on all aspects of the pandemic, to some account making up 4% of the 2020 scientific output indexed by major bibliometric databases. This has led also to scientometric analyses trying to uncover patterns and trends in this vast literature. In this paper we have looked into one aspect which has received little attention so far: the funding of Covid-19 research. It is important to understand how past and current funding has contributed to tackle the pandemic. Such an understanding can not only inform the design of adequate mechanisms for future emergencies but can also help funders meet their accountability obligations. As the pandemic has shown, scientific evidence that generates sound knowledge on which solutions and public policies are built has to compete with well-organised disinformation campaigns. The disclosure of funding sources can also enhance the transparency of the research process and increase the public's confidence in scientific findings.

The main objective of this paper was to explore the extent to which openly available datasets (Crossref funding data) can help in the study of funding of Covid-19 research (operationalised by the CORD-19 dataset). We also aimed to make a comparison with the availability of funding data in proprietary bibliometric databases (Scopus and WoS).

We found that only 17% of the CORD-19 publications with DOIs have funding data in Crossref. This rate was higher for the proprietary databases: 24% for Scopus and 28% for WoS. Considering only publications indexed by a database, we found that 33% of the CORD-19 publications indexed in Scopus have funding data. The corresponding share for WoS is 43%.

In terms of the main funders of Covid-19 research, the three databases paint a broadly similar picture. The three funders with most publications in CORD-19 are the US Department of Health and Human Services (mainly the NIH), the National Natural Science Foundation of China and the European Commission. There are some differences in lower ranks.

By comparing publications with funding data in the three databases, we found a relatively low overlap. Considering only publications indexed in all three databases, only 33% of the publications with funding data in at least one database have funding data in all three databases. For the two proprietary databases, the overlap is 64%.

We also assessed, based on small samples, the accuracy of the funding data present in one database but not in the other two. Our analysis shows that most funding data exclusive to WoS (i.e., data available neither in Crossref nor in Scopus) matched with funding information in the full text of publications. The share of publications for which we could not confirm the correctness of the funding data based on the full text of the publication was also relatively low in Crossref, but it was quite high in Scopus.

After observing that the list of top funders from Scopus includes more pharmaceutical companies than the lists from Crossref and WoS, we also checked manually a sample of 25 publications which had a company among the funding organisations. We found that in most cases this did not correspond to the funding information included in the paper. Rather, it seems to be the result of an error made by the algorithm used to extract and structure funding data. In most cases, the algorithm incorrectly treats the conflict of interest or disclosure section of a paper as a funding statement. As a result, pharmaceutical companies are often presented as funders of the research presented in a paper, while in fact they are funders or collaborators in other activities of the authors.[5]

The main observation from this study is the *limited coverage of funding data* in open data infrastructures. Using Crossref funding data alone allows to paint only a partial picture of who is funding Covid-19 research. In comparison with proprietary databases, the share of CORD-19 publications (with DOI) which have funding data in Crossref is about 7 percentage points lower than in Scopus and about 11 percentage points lower than in WoS. The limited coverage of funding data in Crossref can be explained by differences in the metadata deposited by publishers to Crossref. While for some publishers we have nearly full coverage of funding data, for others the coverage is relatively low, and there are also publishers that do not deposit funding data at all. We also observed that in proprietary databases the coverage by publisher rarely exceeds 75%. Publications without funding data may present research for which the authors did not receive any funding. However, it is also possible that the authors did receive funding and did report this in their publication, but that the funding information was not processed properly due to algorithmic mistakes.

A second observation from this study is the *uncertain quality of funding data*. As already mentioned, in a small random sample that we analysed, we found that most funding data related to pharmaceutical companies is based on algorithmic errors, mainly because extraction algorithms confuse conflict of interest statements with funding statements. This issue affects all databases considered but seems more severe in Scopus.

---

[5] Scopus informed us that it currently focuses on optimising its funding data for the top 300 funders. While this may be good enough for global comparisons, our findings show that it may give an inaccurate picture in more specific analyses.

In the following we offer some reflections on how the availability of funding data can be improved.

Authors provide funding information in their papers in order to comply with requirements of publishers and funders. On the one hand, publishers' ethical guidelines increasingly require disclosure of funding sources. In case of research with commercial or political interests at stake, this transparency helps readers assess the extent to which the credibility of the findings may be related to possible conflicts of interests of the authors. This is standard practice in many medical journals. On the other hand, almost all major research funders require - often as part of their grant terms and conditions - that grant holders explicitly acknowledge the funder's support in publications to which the funding has contributed.

By collecting and making available funding data provided in funding statements in publications, publishers provide an important service to various stakeholders, including:

- to readers by providing transparency on the sponsor of a study.
- to authors by helping them comply with funders' requirements.
- to funders by allowing them to easily identify the results of their funding.
- to scientometricians by enabling them to study the effectiveness and impact of funding practices.

However, as the results presented in this paper show, funding data in open infrastructures is lacking in terms of coverage and to some extent also quality.

Publishers should be encouraged to sustain and intensify their efforts to submit funding data to Crossref. One improvement that could be considered is to extend the data that can be provided to Crossref to also include the raw funding information text. This would allow publishers who cannot commit resources to extract structured funding data from papers to participate in this effort, overcoming the current situation in which publishers need to choose between providing structured funding data to Crossref and providing no funding data at all. Another advantage would also be that having the funding statements as provided in papers could help improve the quality of funding data. As better algorithms become available, they can be applied, also retrospectively, to the available funding statements to turn these statements into high-quality structured funding data.

Funders also have an important role in improving both the availability and the quality of funding data in open data infrastructures. In particular, funders should support efforts of open scholarly infrastructures to create persistent identifiers. The Funder Registry used by Crossref and Scopus and the deposition of funding data to Crossref should be seen as part of broader efforts to improve the availability of high-quality open funding data, building on past efforts - like the guidelines of the UK Research Information Network (RIN, 2008), and continuing to evolve. Crossref recently started an initiative to assign DOIs to research grants[6]. Funders should take up this opportunity and offer guidance to grant holders on how persistent identifiers for grants should be used in funding statements in publications.

As scientific results increasingly impact our daily life, researchers, funders, publishers, research organisations and society at large share an interest in safeguarding the trust and credibility that scholar communication enjoys. Working together on realizing high-quality open funding data is an essential step in this endeavor.

---

[6] https://www.crossref.org/blog/request-for-feedback-on-grant-identifier-metadata/

**Acknowledgments**

We are grateful to Clara Calero, Dan Gibson, and Jeroen van Honk (CWTS), Ginny Hendricks (Crossref), M'hamed El Aisati (Elsevier Scopus), and Gali Halevi (Clarivate Web of Science) for their feedback on an earlier draft of this paper.

**Author contributions**

Alexis-Michel Mugabushaka: Conceptualization; Data curation; Formal analysis; Investigation; Methodology; Visualization; Writing – original draft

Nees Jan van Eck: Conceptualization; Data curation; Formal analysis; Investigation; Methodology; Visualization; Writing – review & editing

Ludo Waltman: Conceptualization; Visualization; Writing – review & editing

**Competing interests**

From 2018 until 2020, Alexis-Michel Mugabushaka was a member of Crossref's Funders Advisory Group on persistent identifiers for grants.

**Funding information**

The authors did not receive any funding for the research reported in this paper.

**Data availability**

The version of the CORD-19 dataset used in this paper was released by Microsoft Academic on 22 February 2021 and corresponds to the release from 15 February 2021 by the Allen Institute. It is accessible at https://magcord19.blob.core.windows.net/mapping/2021-02-22-CORD-19-MappedTo-2021-02-15-MAG-Backfill.csv.

The mapping of funding organisations to the corresponding top-level entities (see Section 2.2.2) is available in Zenodo (Van Eck & Mugabushaka, 2021).
https://doi.org/10.5281/zenodo.5562841

The supplementary material is also available in Zenodo (Mugabushaka et al., 2022).

https://doi.org/10.5281/zenodo.6805409